\documentclass[%
 reprint,
 amsmath,amssymb,
 aps,
]{revtex4-1}
\usepackage{mathrsfs}
\usepackage{times}
\usepackage{graphicx}
\usepackage{dcolumn}
\usepackage{bm}

\begin{document}
\title{Optical switch based on dressed intracavity dark states}

\author{Miaodi Guo}
\author{Xuemei Su}
 \email{suxm@jlu.edu.cn}

\affiliation{%
 Key Lab of Coherent Light, Atomic and Molecular Spectroscopy, Ministry of Education; and College of Physics, Jilin University, Changchun 130012, People's Republic of China}%

\date{\today}
\begin{abstract}
We present a scheme to realize two-direction optical switch by a single-mode optical cavity containing some four-level atoms. The high switching efficiency can be obtained through low photon loss and large third-order nonlinear susceptibility of this N-type atomic system in cavity. Without the microwave source, it can be reduced to a $\Lambda$-type atomic system where a coupling laser is used to realize single intracavity electromagnetically induced transparency (EIT). Namely, the probe field can be transmitted almost totally at resonance. Thus a two-direction optical switch is operated and the state for forward (backward) direction is set as ``open" (``closed"). When microwave source is introduced, dressed splitting of intracavity dark state happens. The probe field is reflected almost completely at resonance and the state of the optical switch at forward and backward directions (transmitted and reflected channels) is shifted as ``closed" and ``open", respectively. Moreover, this scheme is much advantageous to realize splitting of intracavity dark state because weak microwave field ($\Omega_{m}\sim0.1\gamma_{14}$) induces the coupling between intracavity dark state and one sublevel of ground state. While a strong pump laser ($\Omega_{d}\geq\gamma_{14}$) which couples the intracavity dark state with an excited level is applied to realize this splitting in Ref. \href{https://journals.aps.org/pra/abstract/10.1103/PhysRevA.85.013814}{[Phys. Rev. A 85 013814 (2012)]}.
\newline
~~~~~\newline
PACS number(s): 42.65.Pc, 42.50.Gy, 32.80.Qk, 03.65.Sq
\end{abstract}

\maketitle

\section{Introduction}
Optical communications and quantum information networks have attracted more and more attention. As one of the most important elements in all-optical networks, optical switch with high efficiency becomes hot topic these years. Many schemes in this field including interaction-free all-optical switch \cite{OL35/2376,PRA89/023806,PRA79/063830,PRA82/063826,COP15/092702}, optical-bistability switch \cite{PRA93/023806,PRA83/053821,PLA376/2565}, fiber-optical switch \cite{PRL111/193601} and wide-band optical switch based on white light cavity \cite{PRA86/033828,OC313/416,PRA93/043819} have been proposed. The efficient optical switch can be realized when optical nonlinearity is increased and the photon loss of optical system is decreased \cite{PRL64/1107}. It is possible to obtain large optical nonlinearity by effect of quantum interference, such as the powerful quantum coherence technique in $\Lambda$-type EIT or N-type EIT-enhanced atomic system  \cite{OL31/2625,PRA65/053802,OL21/1936,PRL81/3611,OL26/548,PRA68/041801,PRL97/063901} and tunneling induced transparency in nanostructure \cite{APL70/3455,JMO59/729,PRL95/057401,RMP82/2257}. Nowadays, intracavity EIT and modified EIT \cite{OL23/295,OL33/46,PRA87/053802,PRA85/013840,OC358/73} play a vital role in realization of high-efficiency optical switch with the development of cavity quantum electrodynamics (cavity-QED) \cite{Nature465/755,RMP87/1379}, which can be used to suppress the photon loss and enhance the optical nonlinearity.

In this paper we propose a scheme to switch the probe field with different two-direction output and high efficiency in a four-level atom-cavity system. The N-type four-level atomic structure is similar to that of Harris and Yamamoto \cite{PRL81/3611}. But here the switching field couples two sublevels of ground state rather than the levels of a ground state and an excited state, thus the line width in our scheme will be compressed to a great extent. When the four-level atoms are placed in a resonant cavity, smaller power of switching field can create highly efficient optical switch. Besides, with the aid of cavity, we can manipulate the probe field to be output in two directions. One is the output in reflected channel based on interaction free optical switch, the other is controllable double-frequency output in transmitted channel. A two-channel optical switch is proposed in three-level atom-cavity system \cite{PRA89/023806}, but its efficiency is comparatively low because of larger loss from larger decay rate of excited level which is coupled with a ground level by optical field. In our scheme the switching field (microwave field) is applied to couple two sublevels $|2\rangle$ and $|3\rangle$ of ground state and thus lower loss from decay rate of ground state is utilized instead of excited state. Here the principle behind the advantageous optical switch is based on dressed splitting of intracavity dark state \cite{PRA85/013814} rather than interaction free for bright/dark polaritons \cite{PRA89/023806,COP15/092702} or the creation of intracavity dark state \cite{PRA82/033808}. The switching efficiency is as large as $\eta_{T}=0.98$ ($\eta_{R}=0.94$) in transmitted (reflected) channel. We believe optical switch based on this scheme has the potential to be applied in high-efficiency quantum networks and optical communications.

This paper mainly consists of following three parts. In Sec. 2, we perform theoretical analysis based on master equation method and input-output theory \cite{Walls2007}. In Sec. 3, we explore optimal condition on switching efficiency. At last, we exhibit the conclusions in Sec. 4.

\section{Theoretical Model and Analysis}
The interaction diagram of atom-cavity system is shown in Fig. 1(a). The single mode cavity is driven by a probe laser, and the cavity mode interacts with some four-level atoms at a frequency detuning $\Delta_{ac}=\omega_{c}-\omega_{41}$ from $D_{1}$ line of $^{87}$Rb. The frequency detuning between cavity mode and the probe laser is defined as $\Delta_{c}=\omega_{p}-\omega_{c}$. A coupling laser enters into the cavity vertically to its axis driving atomic transition $|2\rangle\rightarrow|4\rangle$ with a frequency detuning $\Delta_{1}=\omega_{1c}-\omega_{42}$. As switching field, the microwave source  injected into the cavity by a microwave horn induces coupling between atomic states $|2\rangle$ and $|3\rangle$ with frequency detuning $\Delta_{m}=\omega_{m}-\omega_{32}$. Two detectors are used to receive transmitted and reflected intracavity light field, respectively. The atomic configurations in bare states and the corresponding dressed states are presented in Fig. 1(b) and Fig. 1(c). Those four states $|1\rangle$, $|2\rangle$, $|3\rangle$, $|4\rangle$ are corresponding to those of $^{87}$Rb, $5S_{1/2}$ $F=1 (m=1)$, $5S_{1/2}$ $F=1 (m=-1)$, $5S_{1/2}$ $F=2 (m=-1)$ and $5 P_{3/2}$ $F=1$, respectively. $\omega_{c}$ and $\omega_{p}$ are angular frequency of cavity mode and probe field. $\Omega_{1}$ ($\omega_{1c}$) and $\Omega_{m}$ ($\omega_{m}$) are Rabi frequency (angular frequency) of coupling field and microwave field, respectively. $\omega_{41}$ ($\omega_{42}$) is the atomic transition frequency of $|1\rangle\leftrightarrow|4\rangle$ ($|2\rangle\leftrightarrow|4\rangle$). When microwave source is switched off, this atom-cavity system exhibits typical single EIT induced by cavity mode and coupling field where the bright state $|B\rangle=\frac{1}{\sqrt{g^{2}N+\Omega_{1}^{2}}}(g\sqrt{N}|1\rangle+\Omega_{1}|2\rangle)$ is still coupled with cavity mode forming two Rabi splitting peaks (bright polaritons) at $\Delta_{p}=\pm\sqrt{g^{2}N+\Omega_{1}^{2}}$ but the dark state $|D\rangle=\frac{1}{\sqrt{g^{2}N+\Omega_{1}^{2}}}(\Omega_{1}|1\rangle-g\sqrt{N}|2\rangle)$ is decoupled to cavity mode as shown in Fig. 1(c). When microwave source is switched on, both bright polaritons and dark state will be affected on the state expression. When microwave source is weak, the new eigenstates for bright polaritons will approximately be $|B_{1}\rangle\approx\frac{1}{\sqrt2}[|4\rangle+\frac{1}{\sqrt{g^{2}N+\Omega_{1}^{2}}}(g\sqrt{N}|1\rangle+\Omega_{1}|2\rangle)]$ and $|B_{2}\rangle\approx\frac{1}{\sqrt2}[|4\rangle-\frac{1}{\sqrt{g^{2}N+\Omega_{1}^{2}}}(g\sqrt{N}|1\rangle+\Omega_{1}|2\rangle)]$ while dark state $|D\rangle$ will split into two dark states $|D_{1}\rangle\approx\frac{1}{\sqrt2}[|3\rangle+\frac{1}{\sqrt{g^{2}N+\Omega_{1}^{2}}}(\Omega_{1}|1\rangle-g\sqrt{N}|2\rangle)]$ and $|D_{2}\rangle\approx\frac{1}{\sqrt2}[|3\rangle-\frac{1}{\sqrt{g^{2}N+\Omega_{1}^{2}}}(\Omega_{1}|1\rangle-g\sqrt{N}|2\rangle)]$. Moreover, the position of these polaritons can be manipulated by microwave source.  In order to explain the difference of dressed splitting of dark state in this paper and in Ref. \cite{PRA85/013814}, we draw level $|3'\rangle$ and the named pump laser ($\Omega_{d}$) in Ref. \cite{PRA85/013814} inside the red dashed rectangle of Fig. 1(c). The detailed discussion is presented in Sec. 3.
\begin{figure*}[!hbt]
\centering
\includegraphics[width=6cm,bb=55 500 330 700]{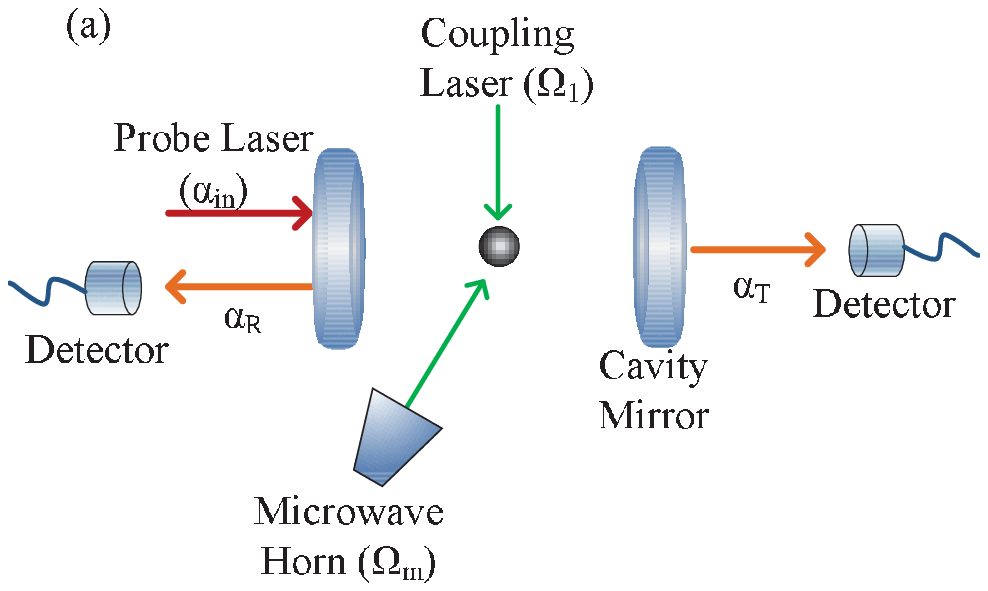}
\includegraphics[width=5cm,bb=40 620 270 820]{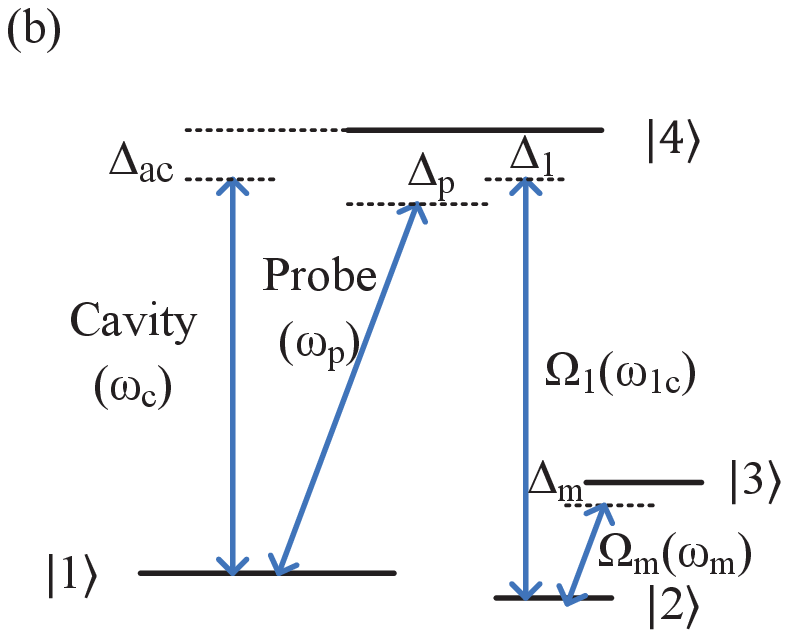}
\includegraphics[width=5cm,bb=40 620 270 820]{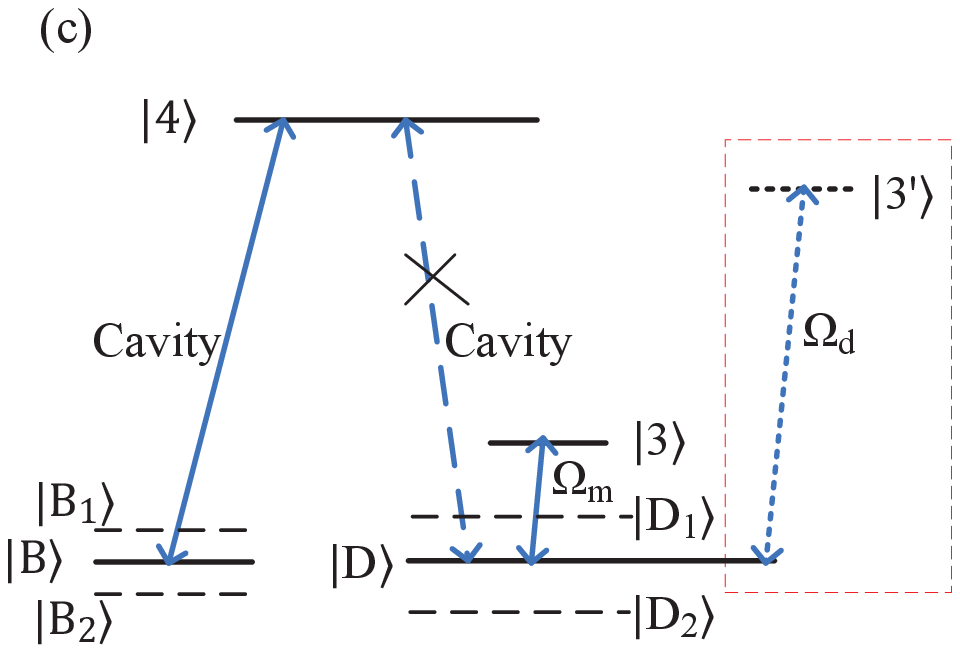}
\caption{Schematic diagram of (a) the atom-cavity system and the energy level structure in atomic (b) bare states and (c) dressed states. A probe laser propagates along the axis of cavity. The coupling field and microwave field are injected into the cavity from outside. In order for a comparison on generation of dark state splitting via ground level $|3\rangle$ and a microwave source $\Omega_{m}$, the excited level $|3'\rangle$ and the named pump laser $\Omega_{d}$ used in Ref. \cite{PRA85/013814} are drawn inside the red dashed rectangle. }
\end{figure*}

For this atom-cavity system, where the cavity mode (probe beam) drives the atomic transition $|1\rangle\rightarrow|4\rangle$, the coupling laser drives the atomic transition $|2\rangle\rightarrow|4\rangle$ and the microwave field couples atomic levels $|2\rangle$ and $|3\rangle$, the Hamiltonian can be written as following,
\begin{equation}
\begin{split}
H=&-\hbar\sum_{j=1}^{N}(\Delta_{p}\sigma_{44}^{j}+(\Delta_{p}-\Delta_{1}+\Delta_{m})\sigma_{33}^{j}+(\Delta_{p}-\Delta_{1})\sigma_{22}^{j})\\
&-\hbar\Delta_{c}a^{\dag}a-\hbar\sum_{j=1}^{N}{(ga^{\dag}\sigma_{14}^{j}e^{i\varphi_{p}}+\Omega_{1}\sigma_{24}^{j}e^{i\varphi_{1}}+\Omega_{m}{\sigma_{23}^{j}}e^{i\varphi_{m}})}\\
&+H.C.,\label{Hamiltonian}
\end{split}
\end{equation}
where the energy nonconserving terms corresponding to the rotating-wave approximation are dropped, $\varphi_{1}$, $\varphi_{m}$ and $\varphi_{p}$ are phases of coupling laser, microwave field and probe laser, respectively, $N$ is the number of atoms in cavity, $\Delta_{p}=\omega_{p}-\omega_{41}$ is the frequency detuning between probe laser and atomic transition $|1\rangle\leftrightarrow|4\rangle$, $\sigma_{mn}^{j}=|m\rangle\langle n|$ ($m, n=1, 2, 3, 4$) is atomic operator for the $j$th atom, $a$ ($a^{\dag}$) is annihilation (creation) operator of the cavity photons and $g=\mu_{14}\sqrt{\omega_{c}/{2\hbar\varepsilon_{0}V}}$ is cavity-atom coupling coefficient.

In general the evolution equation for the expectation value of an operator $\hat{X}$ is written as
\begin{equation}
\frac{\partial\langle\hat{X}\rangle}{\partial{t}}=-\frac{i}{\,\hbar}\langle[\hat{X},\hat{H}]\rangle.\label{evolution}
\end{equation}
Based on the evolution equation above and the relations $\rho_{12}=\langle\sigma_{12}\rangle e^{i(\varphi_{p}-\varphi_{1})}$, $\rho_{13}=\langle\sigma_{13}\rangle e^{i(\varphi_{p}-\varphi_{1}+\varphi_{m})}$, $\rho_{14}=\langle\sigma_{14}\rangle e^{i\varphi_{p}}$, $\rho_{23}=\langle\sigma_{23}\rangle e^{i\varphi_{m}}$, $\rho_{24}=\langle\sigma_{24}\rangle e^{i\varphi_{1}}$, $\rho_{34}=\langle\sigma_{34}\rangle e^{i(\varphi_{1}-\varphi_{m})}$, $\rho_{mm}=\langle\sigma_{mm}\rangle$ and $\rho_{mn}=\rho_{nm}^{*}$, we can derive the following master equations for $\rho_{mn}$ under considering the decay process,
\begin{equation}
\begin{split}
\dot\rho_{11}&=ig(\alpha^{*}\rho_{14}-\alpha\rho_{41})+\frac{\Gamma_{4}}{3}\rho_{44},\\
\dot\rho_{12}&=[i(\Delta_{p}-\Delta_{1})-\gamma_{12}]\rho_{12}+i\Omega_{1}\rho_{14}+i\Omega_{m}\rho_{13}-ig\alpha\rho_{42},\\
\dot\rho_{13}&=[i(\Delta_{p}-\Delta_{1}+\Delta_{m})-\gamma_{13}]\rho_{13}+i\Omega_{m}\rho_{12}-ig\alpha\rho_{43},\\
\dot\rho_{14}&=(i\Delta_{p}-\gamma_{14})\rho_{14}+ig\alpha(\rho_{11}-\rho_{44})+i\Omega_{1}\rho_{12},\\
\dot\rho_{22}&=i\Omega_{1}(\rho_{24}-\rho_{42})+i\Omega_{m}(\rho_{23}-\rho_{32})+\frac{\Gamma_{4}}{3}\rho_{44},\\
\dot\rho_{23}&=(i\Delta_{m}-\gamma_{23})\rho_{23}+i\Omega_{m}(\rho_{22}-\rho_{33})-i\Omega_{1}\rho_{43},\\
\dot\rho_{24}&=(i\Delta_{1}-\gamma_{24})\rho_{24}+ig\alpha\rho_{21}+i\Omega_{1}(\rho_{22}-\rho_{44})-i\Omega_{m}\rho_{34},\\
\dot\rho_{33}&=i\Omega_{m}(\rho_{32}-\rho_{23})+\frac{\Gamma_{4}}{3}\rho_{44},\\
\dot\rho_{34}&=[i(\Delta_{1}-\Delta_{m})-\gamma_{34}]\rho_{34}+ig\alpha\rho_{31}+i\Omega_{1}\rho_{32}-i\Omega_{m}\rho_{24},\\
\dot\rho_{44}&=ig(\alpha\rho_{41}-\alpha^{*}\rho_{14})+i\Omega_{1}(\rho_{42}-\rho_{24})-\Gamma_{4}\rho_{44},\label{density}
\end{split}
\end{equation}
where $\alpha=\langle{a}\rangle$ and $\alpha^{*}=\langle{a^{\dag}}\rangle$ \cite{PRA93/023806}. $\Gamma_{4}$ is the spontaneous emission decay rate from excited state $|4\rangle$ (here we assume the spontaneous emission from state $|4\rangle$ to state $|1\rangle$, $|2\rangle$ or $|3\rangle$ is identical) and $\gamma_{ij}=(\Gamma_{i}+\Gamma_{j})/2$ is the decoherence rate between states $|i\rangle$ and $|j\rangle$, in which $\Gamma_{1}=\Gamma_{2}=\Gamma_{3}=0$. We assume that the population of atomic levels $|2\rangle$, $|3\rangle$ and $|4\rangle$ can be ignored in the initial time and $\rho_{11}+\rho_{22}+\rho_{33}+\rho_{44}=1$.

In our system, we consider a Fabry-Perot cavity with field decay rates $\kappa_{l}$ and $\kappa_{r}$ through the left and right mirrors respectively and the total field decay rate $\kappa$. The operator $a_{in,l}$ ($a_{in,r}$) is the field operator of an electromagnetic mode impinging from the left (right) side and $a_{out,l}$ ($a_{out,r}$) describes the outgoing field to the left (right) side. The output field satisfy the following equations \cite{Walls2007},
\begin{equation}
\begin{split}
a_{out,l}+a_{in,l}=\sqrt{2\kappa_{l}\tau}\,a,\\
a_{out,r}+a_{in,r}=\sqrt{2\kappa_{r}\tau}\,a,\label{input-output}
\end{split}
\end{equation}
where $\kappa_{i}=T_{i}/2\tau(i=l, r)$ ($T_{i}$ is the mirror transmission and $\tau$ is the photon round-trip time inside the cavity) \cite{PRA87/053802} and $\langle a_{in,r}\rangle=0$ in our scheme. The intracavity field operator $a$ can be solved by Heisenberg-Langevin motion equation following,
\begin{equation}
\langle\dot{a}\rangle=-\frac{i}{\hbar}\langle[a,H]\rangle-\kappa\langle{a}\rangle+\sqrt{2\kappa_{l}/\tau}\langle{a_{in}}\rangle.\label{motioneq1}
\end{equation}
Inserting $H$ into Eq. (\ref{motioneq1}), we have
\begin{equation}
\langle\dot{a}\rangle=i\Delta_{c}\langle a\rangle+igN\rho_{14}-\kappa \langle a\rangle+\sqrt{2\kappa_{l}/\tau}\langle a_{in}\rangle.\label{motioneq2}
\end{equation}

By solving equations (\ref{density}) and (\ref{motioneq2}) in steady state ($\dot\rho_{mn}=0$ and $\langle\dot{a}\rangle=0$), the intracavity field $\alpha$ can be derived as
\begin{equation}
\alpha=\frac{\sqrt{2\kappa_{l}/\tau}\alpha_{in}}{\kappa-i\Delta_{c}-i\chi},\label{intracavity}
\end{equation}
where $\chi=-g^{2}N(\Delta_{p}^{2}-\Omega_{m}^{2})/[i\gamma_{14}(\Delta_{p}^{2}-\Omega_{m}^{2})+\Delta_{p}(\Delta_{p}^{2}-\Omega_{1}^{2}-\Omega_{m}^{2})]$.
Here we ignore $|\alpha|^{2}$ and $o(\alpha)$, the high order infinitesimal of $\alpha$ since probe field is much weaker than the coupling field.
The amplitude transmission ($t_{out,r}$) and reflection ($r_{out,l}$) coefficients are derived from Eqs. (\ref{input-output}) and (\ref{intracavity}) as following,
\begin{equation}
\begin{split}
t_{out,r}&=\alpha_{out,r}/\alpha_{in}=\frac{2\sqrt{\kappa_{l}\kappa_{r}}}{\kappa-i\Delta_{c}-i\chi},\\
r_{out,l}&=\alpha_{out,l}/\alpha_{in}=1-\frac{2\kappa_{l}}{\kappa-i\Delta_{c}-i\chi}.\label{output}
\end{split}
\end{equation}
For a symmetrical cavity with $\kappa_{l}=\kappa_{r}=\kappa/2$, according to the definition of the output field intensities $I_{T}=|t_{out,r}|^{2}I_{in} $ and $I_{R}=|r_{out,l}|^{2}I_{in}$ for transmitted and reflected channels respectively, we have transmission (T) and reflection (R) as following,
\begin{equation}
\begin{split}
T&=I_{T}/I_{in}=|t_{out,r}|^{2}\\
&=\frac{\kappa^{2}(A^{2}+B^{2})}{C^{2}+2C(A\kappa-B\Delta_{c})+(A^{2}+B^{2})(\Delta_{c}^{2}+\kappa^{2})},\\
R&=I_{R}/I_{in}=|r_{out,l}|^{2}\\
&=\frac{C^{2}-2BC\Delta_{c}+(A^{2}+B^{2})\Delta_{c}^{2}}{C^{2}+2C(A\kappa-B\Delta_{c})+(A^{2}+B^{2})(\kappa^{2}+\Delta_{c}^{2})},\label{TR}
\end{split}
\end{equation}
where $A=\gamma_{14}(\Delta_{p}^{2}-\Omega_{m}^{2})$, $B=\Delta_{p}(\Delta_{p}^{2}-\Omega_{1}^{2}-\Omega_{m}^{2})$, $C=(g\sqrt{N})^{2}(\Delta_{p}^{2}-\Omega_{m}^{2})$ and $I_{in}$ is the intensity of the input field.

\section{Results and Discussion}

\subsection*{\emph{3.1. Dressed splitting of intracavity dark state}}
\begin{figure}[!hbt]
\centering
\includegraphics[width=4.2cm]{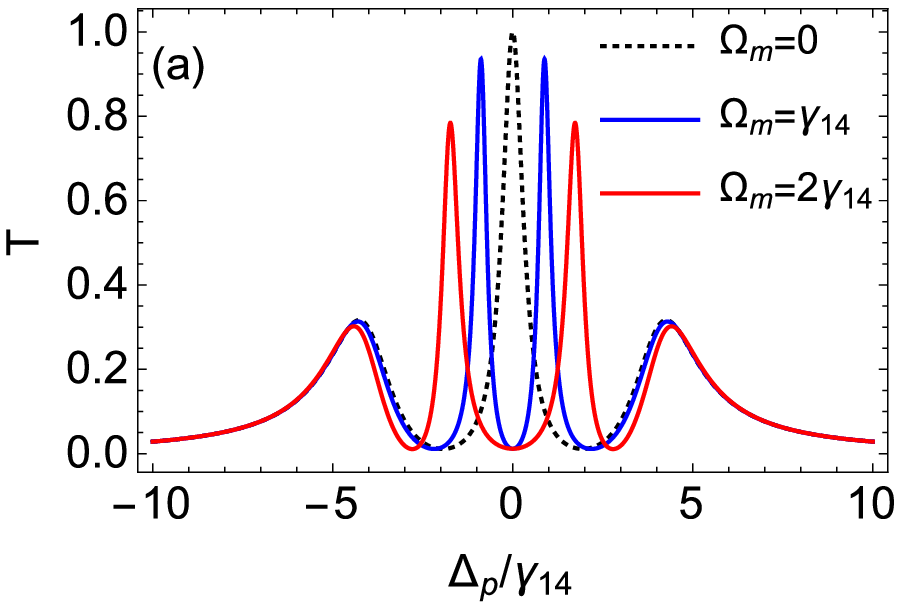}
\includegraphics[width=4.2cm]{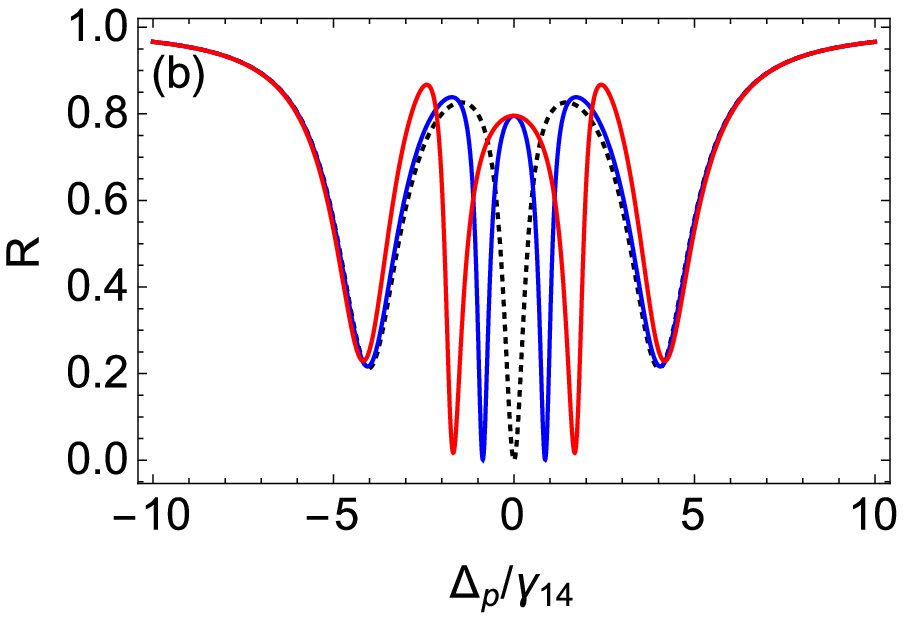}
\caption{The transmission (a) and reflection (b) of intracavity probe field. The parameters are $g\sqrt{N}=3.5\gamma_{14}$, $\kappa=1.5\gamma_{14}$, $\Omega_{1}=2\gamma_{14}$ and $\Omega_{m}=0$, $\gamma_{14}$, $2\gamma_{14}$ for black dotted lines, blue solid lines and red solid lines respectively.}
\end{figure}
We show dressed splitting of intracavity dark state by changing the strength of microwave field in Fig. 2. When $\Omega_{m}=0$, the output field exhibits a narrow peak at $\Delta_{p}=0$ and two broad peaks located on its left and right sides at $\Delta_{p}=\pm\sqrt{g^{2}N+\Omega_{1}^{2}}$ as shown in Fig. 2(a) and 2(b) with black dotted curves. The middle EIT peak reperents the intracavity dark state $|D\rangle$ which is uncoupled with cavity mode therefore the line width of EIT peak will not be influenced by cavity mode decay, hence it is narrower than two sidebands. Two broad peaks are resulted from Rabi splitting of bright state $|B\rangle$. However, when we apply microwave source, single EIT peak splits into two peaks near resonance frequency ($\Delta_{p}\approx\pm{g\sqrt{N}\Omega_{m}}/{\sqrt{g^{2}N+\Omega_{1}^{2}+\Omega_{m}^{2}}}$), which is induced by atomic coherence between states $|D\rangle$ and $|3\rangle$ caused by microwave field. These dark polaritons $|D_{1}\rangle$ and $|D_{2}\rangle$ do not contain excited state so that they are scarcely affected by the spontaneous emission, therefore their line width is much narrower than that of bright polaritons $|B_{1}\rangle$ and $|B_{2}\rangle$. While the microwave field takes little effect on the bright polaritons for it is so weak that the impact can be neglected. Without approximation, we obtain the position for two bright polaritons ($P_{B}$) and two dark polaritons ($P_{D}$) under resonance condition ($\Delta_{p}=\Delta_{1}=\Delta_{m}=0$) by solving Hamiltonian matrix,
\begin{equation}
\begin{split}
P_{B}\!=\!\pm\sqrt{\frac{g^{2}N\!+\!\Omega_{1}^{2}\!+\!\Omega_{m}^{2}\!+\!\sqrt{(g^{2}N\!+\!\Omega_{1}^{2}\!+\!\Omega_{m}^{2})^{2}\!-\!4g^{2}N\Omega_{m}^{2}}}{2}},\\
P_{D}\!=\!\pm\sqrt{\frac{g^{2}N\!+\!\Omega_{1}^{2}\!+\!\Omega_{m}^{2}\!-\!\sqrt{(g^{2}N\!+\!\Omega_{1}^{2}\!+\!\Omega_{m}^{2})^{2}\!-\!4g^{2}N\Omega_{m}^{2}}}{2}}.
\end{split}
\end{equation}

\begin{figure}[!hbt]
\centering
\includegraphics[width=6cm]{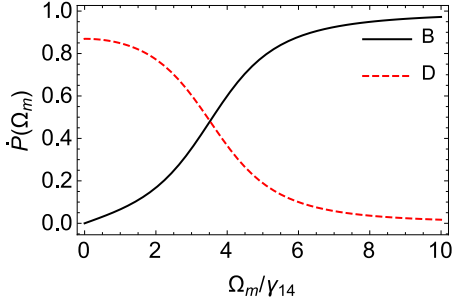}
\caption{The slope of position varying curves $P(\Omega_{m})$ versus $\Omega_{m}$. $g\sqrt{N}=3.5\gamma_{14}$, $\kappa=1.5\gamma_{14}$ and $\Omega_{1}=2\gamma_{14}$. The black solid line and red dashed line are corresponding to bright polaritons and dark polaritons, respectively. }
\end{figure}
To explore how microwave source affects the position of these polaritons, we draw a picture for $\dot P_{B}(\Omega_{m})$ and $\dot P_{D}(\Omega_{m})$ in Fig. 3. As we can see, if microwave field is weak enough, e.g. $\Omega_{m}=\gamma_{14}$, $\dot P_{B}\approx0.06$ for bright polaritons which is ten times smaller than $\dot P _{D}\approx0.84$ for dark polaritons. This is why different position changes via the same $\Omega_{m}$ between bright polaritons and dark polaritons are shown in Fig. 2. It shows in Fig. 3 that the position changes for bright polaritons become more obvious when $\Omega_{m}>2\gamma_{14}$ while the position changes for dark polaritons are smaller and smaller with increasing $\Omega_{m}$. When $\Omega_{m}$ is up to $6\gamma_{14}$, the position for dark polaritons keeps almost changeless by increasing $\Omega_{m}$. Therefore, according to Fig. 2 and Fig. 3, we propose an optical switch with frequency manipulated and direction controllable where we take no account of the output for bright polaritons. By switching on microwave source, transmitted channel in resonance frequency $\Delta_{p}=0$ is closed while reflected channel in resonance frequency is open, at the same time the transmitted channel with double frequency is open. Moreover, based above discussion, the double frequency output for transmitted channel can be adjusted by manipulating $\Omega_{m}$. So this atom-cavity system can be used to make an efficient optical switch based on dressed splitting of intracavity dark state by a weak microwave field. The detailed discussion will be presented in Sec. 3.2.

\begin{figure*}[!hbt]
\centering
\includegraphics[width=5.5cm]{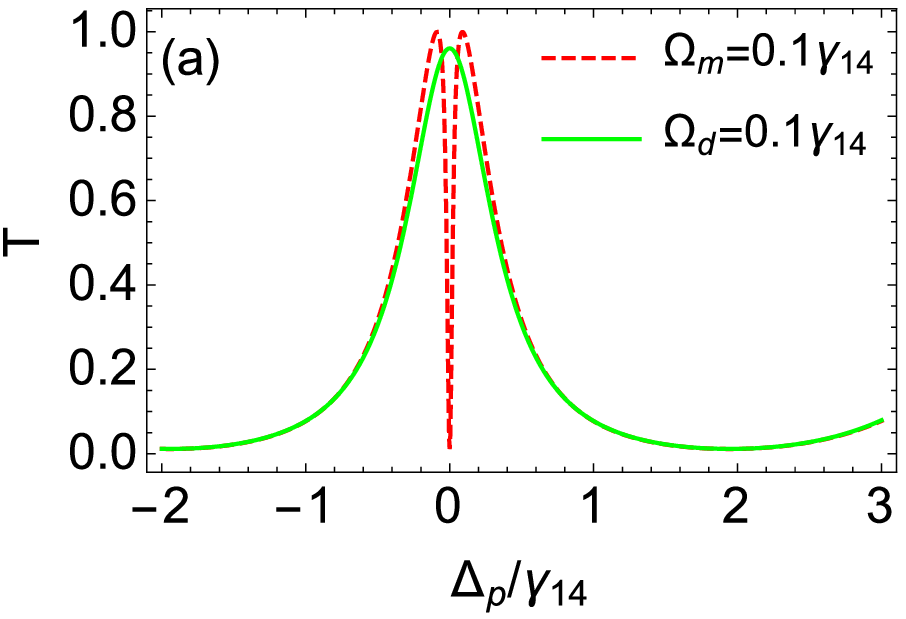}
\includegraphics[width=5.5cm]{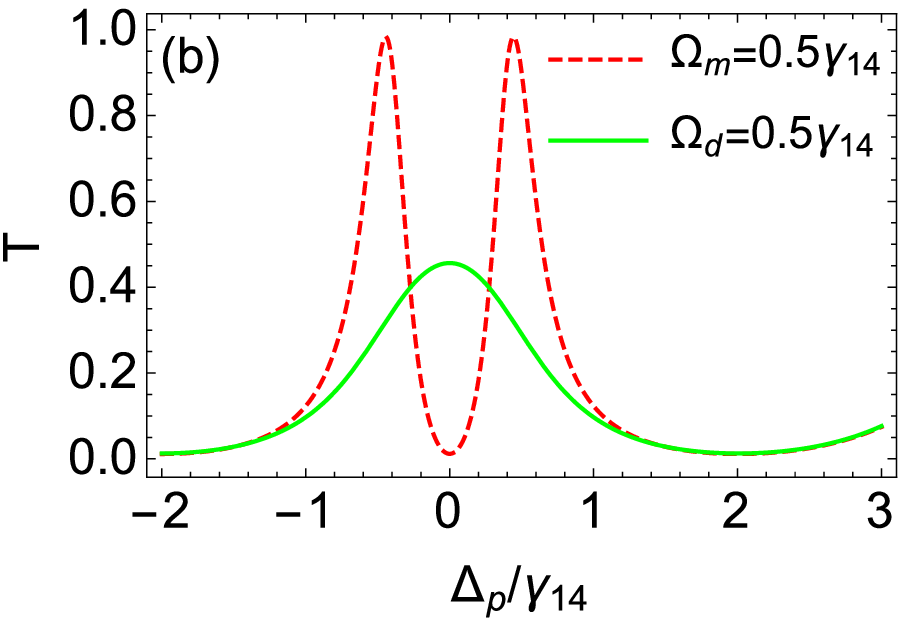}
\includegraphics[width=5.5cm]{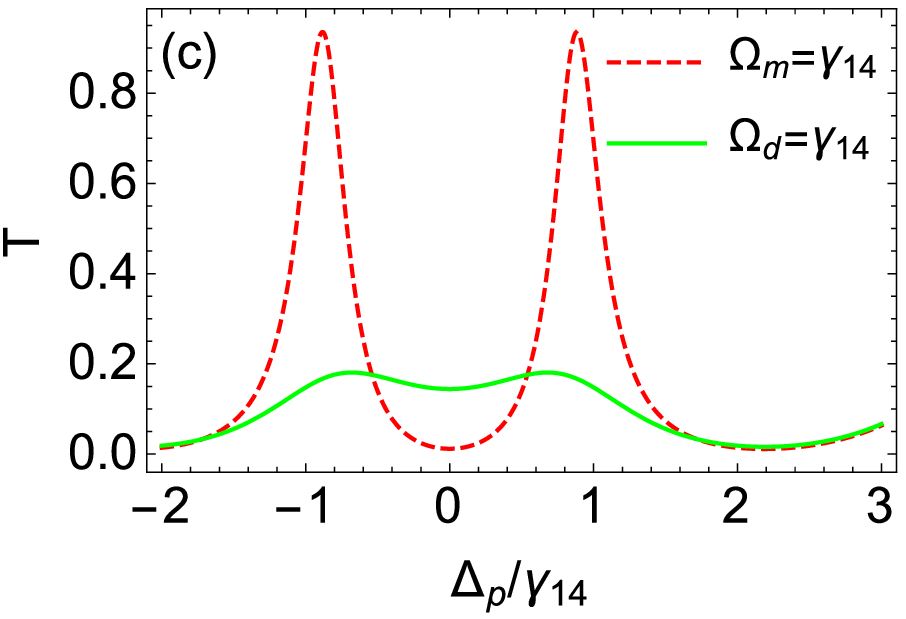}
\caption{The comparison of transmission spectra of dressed intracavity dark states in our scheme (red dashed lines) with the scheme of Yanhua Wang et al in Ref. \cite{PRA85/013814} (green solid lines). Parameters in red lines is similar with that in Fig. 2 and parameters in green lines are referenced from Ref. \cite{PRA85/013814}.}
\end{figure*}
In order to show difference between the dressed splitting of intracavity dark state in the scheme here and in Ref. \cite{PRA85/013814}, we draw transmission spectra around intracavity EIT window in Fig. 4 by different Rabi frequency of dressed field (microwave source here and pump laser in their work). The red dashed lines and green solid lines represent the results in this scheme and in Ref. \cite{PRA85/013814}, respectively. In Fig. 4(a), when Rabi frequency of dressed field is 0.1$\gamma_{14}$, a tiny peak splitting emerges in red curve while no splitting in green solid line. In Fig. 4(b), when Rabi frequency is $0.5\gamma_{14}$, peak splitting is increased in red but no splitting in green line yet. However the transmittivity in green drops to about 0.5 at $\Delta_{p}=0$. In Fig. 4(c), with increasing Rabi frequency up to $\gamma_{14}$, both red dashed line and green solid line exhibit dressed splitting of intracavity EIT. By contrast, the transmittivity at transparency window is still large and it is very close to zero at $\Delta_{p}=0$ in our scheme. This means dressed intracavity dark states appear more easily by coupling with ground state than with an excited state. The reason is obvious that dressed splitting can happen if the Rabi frequency of dressed field is as large as or more than the decay rate of the dressed states. Here except small decay rate of the intracavity dark state, the  decay rate of the ground level $|3\rangle$ is much smaller than that of excited level $|3'\rangle$. According to Ref. \cite{PRA85/013814}, their experimental results are in good agreement with the theoretical predictions. Based on the comparison between our theoretical calculations and theirs, we believe our scheme has more advantages to observe dressed intracavity dark states and can be acted as a high-efficiency optical switch in reality.

\subsection*{\emph{3.2. Switching efficiency in  weak microwave source}}
The switching efficiency $\eta$ is defined as \cite{PRA89/023806}
\begin{equation}
\eta=[I_{out}(\text{open})-I_{out}(\text{closed})]/I_{in}\label{efficiency},
\end{equation}
where $I_{out}(\text{open})$ and $I_{out}(\text{closed})$ represent intensity of output field when optical switch is working at ``open" and ``closed" state, respectively. In order to obtain optimal switching efficiency, we explore the influence from the variables $g\sqrt{N}$, $\Omega_{1}$ and $\Omega_{m}$ and the results are shown in Figs. 5-7.

From Fig. 2(a) and Fig. 2(b), it infers that for transmitted intracavity  field near resonance, $I_{T}\neq0$ when $\Omega_{m}=0$ and $I_{T}=0$ when $\Omega_{m}\neq0$. Therefore, we define $I_{T}(\text{O})$ ($\Omega_{m}=0$) as open state of the switch while $I_{T}(\text{C})$ ($\Omega_{m}\neq0$) as closed state of the switch for transmitted signal. Meanwhile for the reflected intracavity field, $I_{R}=0$ for $\Omega_{m}=0$ while $I_{R}\neq0$ for $\Omega_{m}\neq0$. Hence we choose $I_{R}(\text{O})$ ($\Omega_{m}\neq0$) and $I_{R}(\text{C})$ ($\Omega_{m}=0$) as the open state and closed state of the switch for reflected signal, respectively. The switching efficiency $\eta_{T}$ for transmitted signal and $\eta_{R}$ for reflected signal are obtained as,
\begin{equation}
\begin{split}
\eta_{T}&=\frac{I_{T}(\text{O})-I_{T}(\text{C})}{I_{in}}=\frac{I_{T}(\Omega_{m}=0)-I_{T}(\Omega_{m}\neq0)}{I_{in}},\\
\eta_{R}&=\frac{I_{R}(\text{O})-I_{R}(\text{C})}{I_{in}}=\frac{I_{R}(\Omega_{m}\neq0)-I_{R}(\Omega_{m}=0)}{I_{in}}.
\end{split}
\end{equation}

\begin{figure}[!hbt]
\centering
\includegraphics[width=4.2cm]{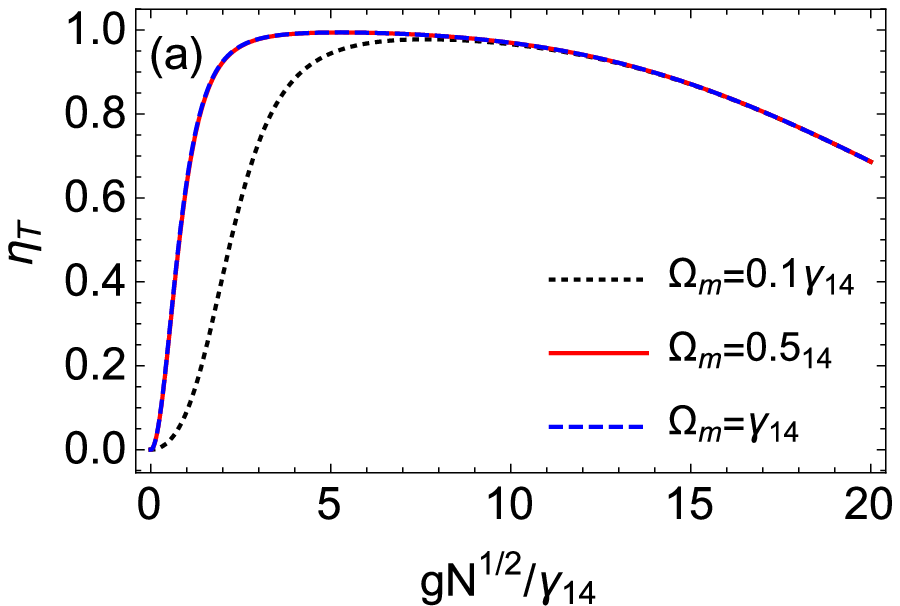}
\includegraphics[width=4.2cm]{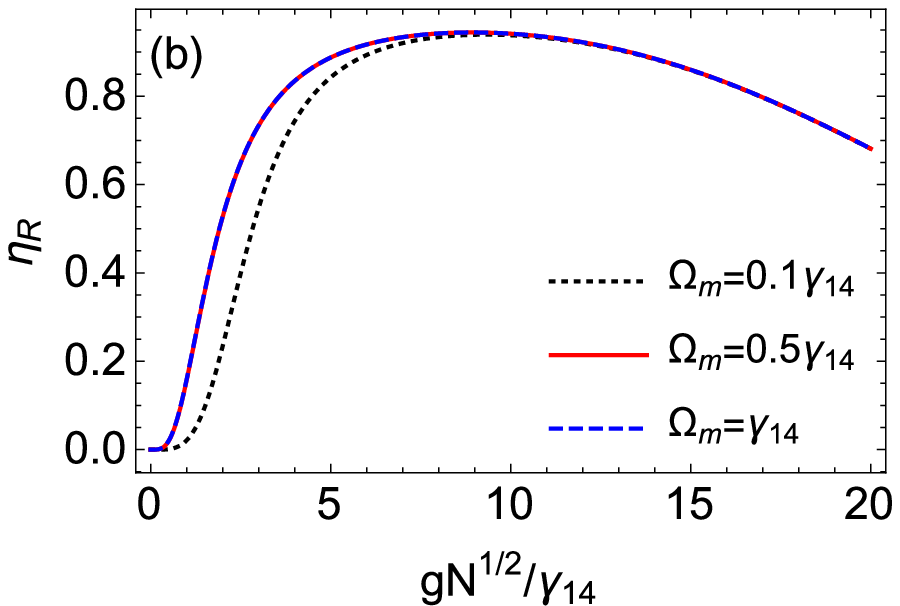}
\caption{Switching efficiency (a) $\eta_{T}$ of transmitted signal and (b) $\eta_{R}$ of reflected signal versus the collective coupling coefficient $g\sqrt{N}$. The parameters are $\Delta_{p}=0.01\gamma_{14}$, $\kappa=1.5\gamma_{14}$,  $\Omega_{1}=2\gamma_{14}$ and $\Omega_{m}=0.1\gamma_{14}$, 0.5$\gamma_{14}$, $\gamma_{14}$ for black dotted lines, red solid lines and blue dashed lines respectively. }
\end{figure}
Fig. 5 shows that with increasing collective coupling coefficient $g\sqrt{N}$, the switching efficiency for both transmitted and reflected signal are being enhanced until they reach a steady optimal value and then start being dropped. In the conditions that $\Omega_{m}=0.5\gamma_{14}$ or $\gamma_{14}$, the switching efficiency tends to be steady when $g\sqrt{N}$ is up to $\sim3\gamma_{14}$ for transmitted signal ($\sim5\gamma_{14}$ for reflected signal). When $g\sqrt{N}$ exceeds $\sim10\gamma_{14}$, the switching efficiency decreases with increasing $g\sqrt{N}$ (for both transmitted and reflected signal). The optimal extent of collective coupling coefficient $g\sqrt{N}$ is $3\gamma_{14}\sim10\gamma_{14}$ for transmitted channel and $5\gamma_{14}\sim12\gamma_{14}$ for reflected channel when $\Omega_{m}=0.5\gamma_{14}$ and $\gamma_{14}$. Similarly we obtained the optimal extent as $5\gamma_{14}\sim10\gamma_{14}$ for transmitted channel and $6\gamma_{14}\sim12\gamma_{14}$ for reflected channel when $\Omega_{m}=0.1\gamma_{14}$.

Next, we explore how coupling laser influences the switching efficiency. We choose collective coupling coefficient $g\sqrt{N}=8\gamma_{14}$.
\begin{figure}[!hbt]
\centering
\includegraphics[width=4.2cm]{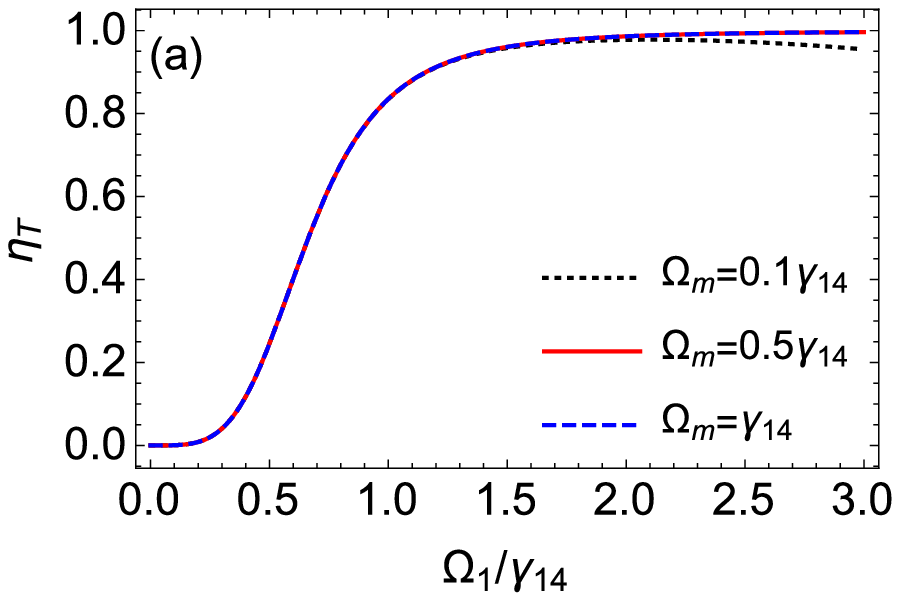}
\includegraphics[width=4.2cm]{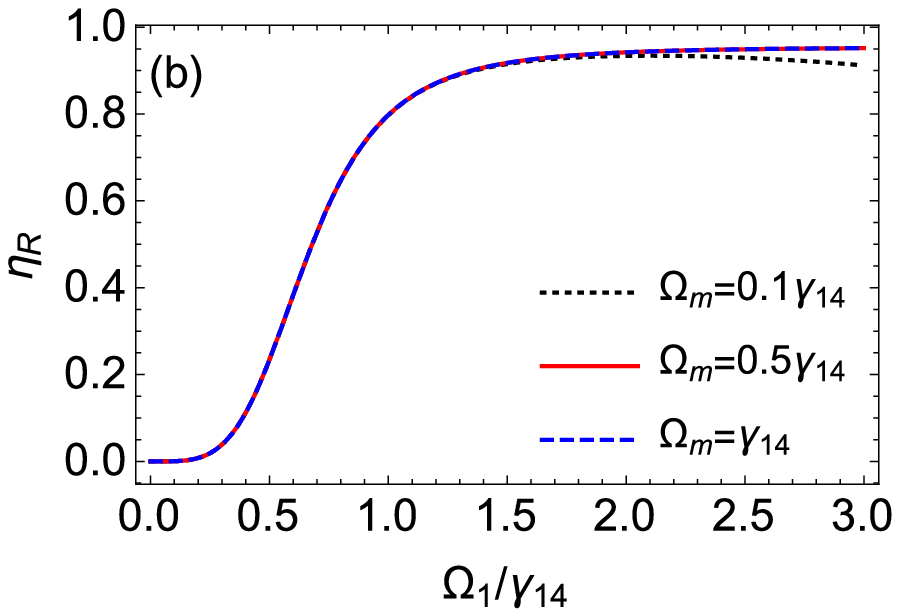}
\caption{The switching efficiency (a) $\eta_{T}$ (transmitted signal) and (b) $\eta_{R}$ (reflected signal) versus the Rabi frequency of control laser 1. The parameters here are $\Delta_{p}=0.01\gamma_{14}$, $\kappa=1.5\gamma_{14}$ and $g\sqrt{N}=8\gamma_{14}$. The black dotted lines, red solid lines and blue dashed lines are corresponding to $\Omega_{m}=0.1\gamma_{14}$, $0.5\gamma_{14}$ and $\gamma_{14}$ respectively. }
\end{figure}
Fig. 6 shows that when $\Omega_{1}$ is less than $0.5\gamma_{14}$, the switching efficiency for both transmitted and reflected channel are about zero. When we increase $\Omega_{1}$ there is a sharp slope on the curves at first, then a stable tendency appears when $\Omega_{1}\geq1.5\gamma_{14}$. As shown, for a given $\Omega_{1}$, $\Omega_{m}$ changing from $0.1\gamma_{14}$ to $\gamma_{14}$ has little effect on $\eta_{T}$ or $\eta_{R}$. It is for the reason that the perturbation of $|3\rangle$ with $\Omega_{m}$ can cause the splitting of dark state easily while has small effect on the radition rate of excited state.

In the following, we focus on the relation between the microwave source (switching field) and switching efficiency by choosing appropriate values of $g\sqrt{N}$ and $\Omega_{1}$. As discussed in introduction, a three-level atom-cavity system can be another type of all-optical switch \cite{PRA82/033808} which is equivalent to the four-level atom-cavity system here without the microwave source, namely the switching field is the coupling field $\Omega_{1}$. For EIT optical switch in Ref. \cite{PRA82/033808}, the signal field and the switching field interact simultaneously with the atoms and are directively coupled. However in our scheme, there is no transmission but reflection at $\Delta_{p}=0$ when microwave source is applied, therefore the direct coupling between probe field and switching field does not exist. Hence, the optical switch is performed interaction free in reflected channel. What's more, for transmitted channel, double-frequency output with controllable frequency interval can be realized by manipulating $\Omega_{m}$. In order to compare the switching efficiency of those two types of optical switch, we plot the switching efficiency versus Rabi frequency of the switching field in Fig. 7. The optical switch in four-level atom-cavity system is based on dressed splitting of intracavity dark state while it is based on the generation of intracavity dark state in three-level atom-cavity system. Switching efficiency of the three-level atom-cavity system is expressed as $\eta_{t}$ ($\eta_{r}$) for transmitted channel (reflected channel). According to the definition of Eq. (\ref{efficiency}), $\eta_{t}=[I_{T}(\Omega_{m}=0, \Omega_{1}\neq0)-I_{T}(\Omega_{m}=0, \Omega_{1}=0)]/I_{in}$, $\eta_{r}=[I_{R}(\Omega_{m}=0, \Omega_{1}=0)-I_{R}(\Omega_{m}=0, \Omega_{1}\neq0)]/I_{in}$.

As presented in Fig. 7(a) and 7(b), the switching efficiency $\eta_{T}$ and $\eta_{R}$ reach a high stable value when $\Omega_{m}$ is as small as $0.1\gamma_{14}$. While in Fig. 7(c) and 7(d), the switching efficiency $\eta_{t}$ and $\eta_{r}$ do not reach up to 0.8 until $\Omega_{1}=\gamma_{14}$, moreover the efficiency can not reach a stable value until Rabi frequency of the switching field $\Omega_{1}=1.5\gamma_{14}$, which is 15 times as large as that in four-level atom-cavity system. This can be explained for the reason of different decay rate of upper level connected by switching field.In our scheme, the switching field $\Omega_{m}$ couples two ground levels, which means the perturbation caused by microwave source can destroy the formation of intracavity dark state easily and the N-type atomic structure will induce large third-order nonlinear susceptibility, therefore the phenomenon whether incident light goes transparently or not is very sensitive to the microwave source. However, in Ref. \cite{PRA82/033808}, the switching field drives the transition of one ground level and the excited level, thus the switching intensity must be large enough to resist the large decay rate of excited state so that the light can go through the medium completely transparently. That means optical switch based on dressed intracavity dark states instead of single intracavity dark state can be carried out with a much weaker switching field. For transmitted channel, we can choose moderate parameters mentioned above with $\Omega_{m}=0.2\gamma_{14}$ (weak microwave source) to make the switching efficiency $\eta_{T}\approx0.98$. For reflected channel, considering with the absorption of atomic system, the optimal switching efficiency $\eta_{R}\approx0.94$ can be obtained under the same parameters with transmitted channel.
\begin{figure}[!hbt]
\centering
\includegraphics[width=4.2cm]{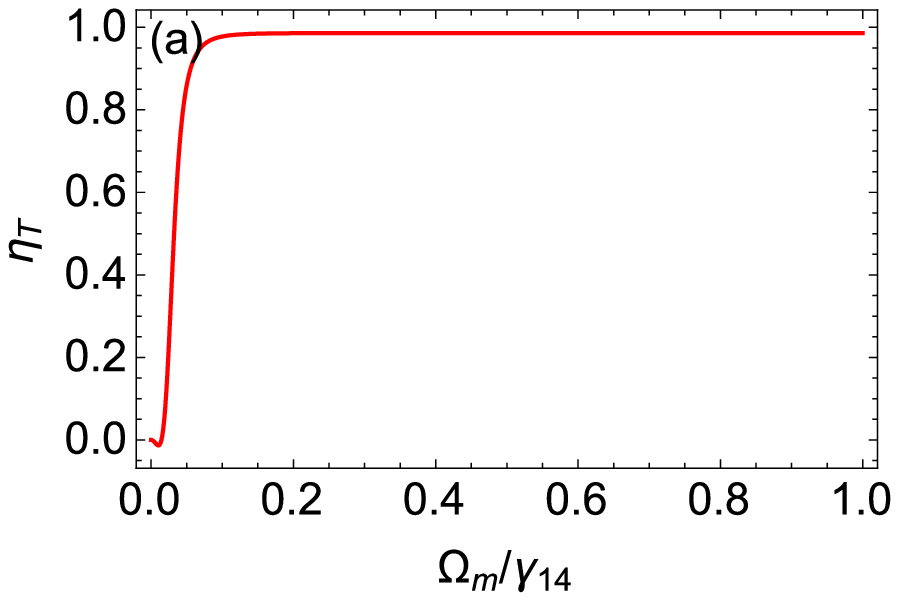}
\includegraphics[width=4.2cm]{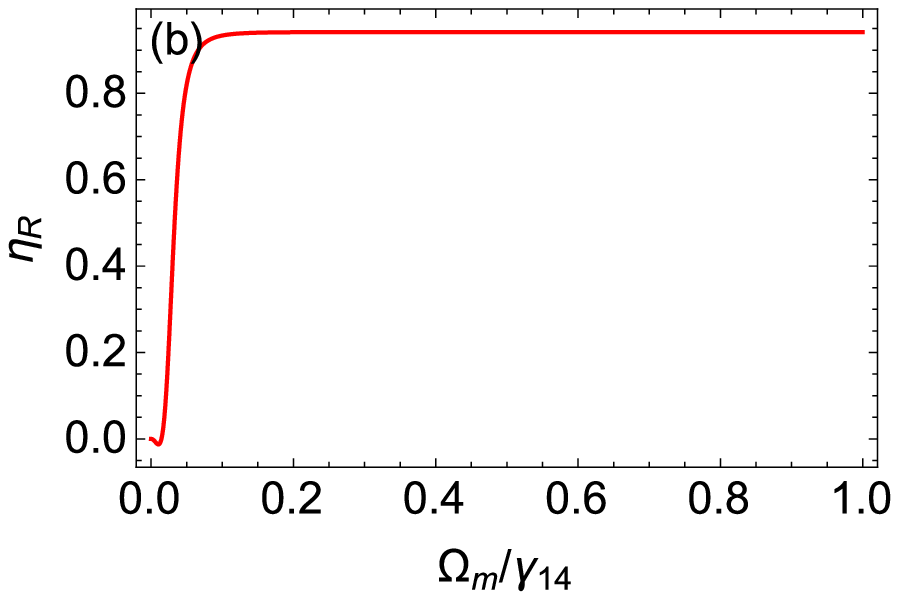}
\includegraphics[width=4.2cm]{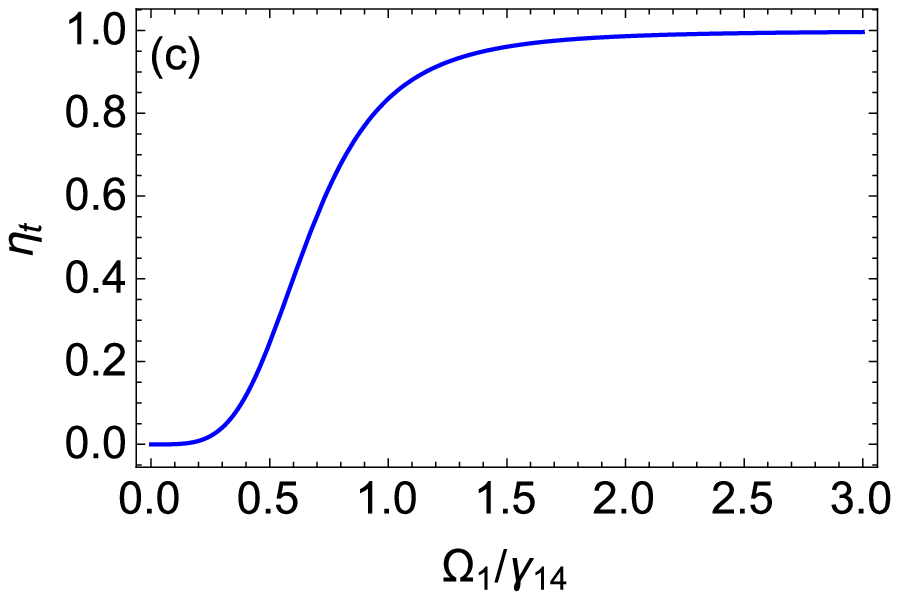}
\includegraphics[width=4.2cm]{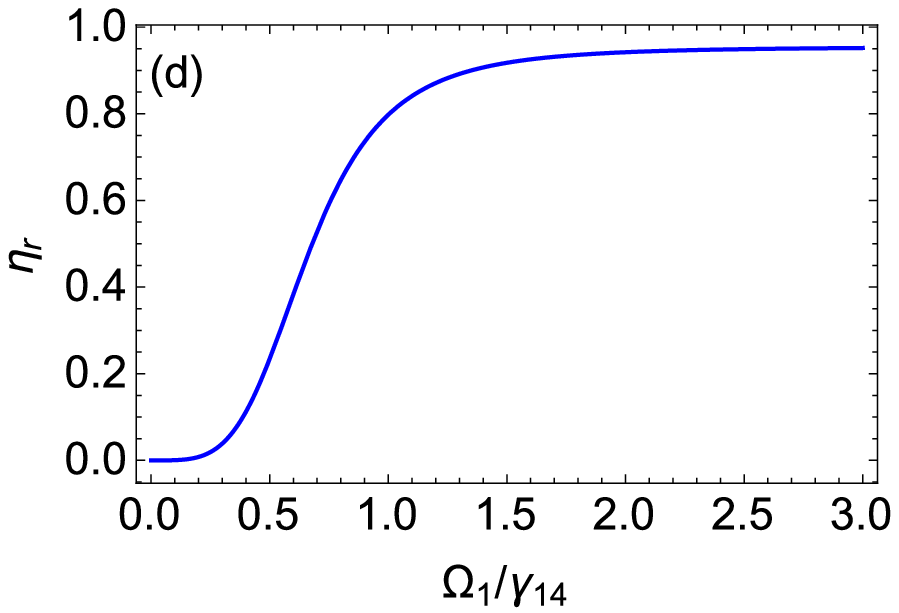}
\caption{The switching efficiency of (a),(c) transmitted channel and (b),(d) reflected channel versus the strength of switching field. The parameters used here are $\Delta_{p}=0.01\gamma_{14}$, $\kappa=1.5\gamma_{14}$, $g\sqrt{N}=8\gamma_{14}$, $\Omega_{1}=2\gamma_{14}$ in (a), (b) and $\Omega_{m}=0$ in (c), (d). }
\end{figure}

\section{Conclusions}
In conclusion, we propose and demonstrate a scheme for realization of two-direction high-efficiency optical switch, which can be performed interaction free at $\Delta_{p}=0$ for reflected channel meanwhile can be used for a double-frequency output with controllable frequency interval for transmitted channel. This is different from Ref. \cite{COP15/092702}, since the optical switch here is based on dressed splitting of intracavity dark state by a weak microwave field coupled to two sublevels of ground state and the switching efficiency will be up to a high stable value via small switching intensity, by which sensitive optical switch  can be performed. Comparing with similar system by using a pump laser to couple a sublevel of ground state with an excited level, it is much easier to induce dressed splitting of intracavity dark state even by a so weak switching field that its Rabi frequency is equal to $0.1\gamma_{14}$ in our scheme. We also discuss the difference of switching efficiency for two types of two-direction optical switches between ours and the three-level atom-cavity system in Ref. \cite{PRA82/033808}. As a result, we find the optical switch in our scheme possesses higher switching efficiency than that based on creation of intracavity dark state. The efficiency of optical switch here can reach $\eta_{T}=0.98$ and $\eta_{R}=0.94$  when microwave source is taken as switching field. Based on the comparison of numerical calculations with previous works by others, our scheme is convinced with more advantages in reality to be applied in the subtle and high-efficiency optical switch.

\section*{Acknowledgment}
We acknowledge financial support by the National Natural Science Foundation of China under Grant No.11174109.

\end{document}